\documentclass[journal=jpccck,manuscript=article]{achemso}

\usepackage{chemformula} 
\usepackage[T1]{fontenc} 

\author{Shingo Urata}
\affiliation[AGC]
{Innovative Technology Laboratories, AGC Inc., 
1-1 Suehiro-cho, Tsurumi-ku, Yokohama, Kanagawa, 230-0045, Japan}
\email{shingo.urata@agc.com}

\title{Modeling Short-Range and Three-Membered Ring Structures in Lithium Borosilicate Glasses using Machine Learning Potential}

\abbreviations{ }
\keywords{Bolosilicate glass, Molecular dynamics, Force field, Machine-learning}

\begin{document}

\newpage

\begin{abstract}
Lithium borosilicate (LBS) glass is a prototypical lithium-ion conducting oxide glass available for 
an all-solid state buttery.
Nevertheless, the atomistic modeling of LBS glass using $ab$ $initio$ (AIMD) and classical molecular dynamics (CMD) simulations 
have critical limitations due to 
computational cost and inaccuracy in reproducing the glass microstructures, respectively.
To overcome these difficulties, a machine-learning potential (MLP) was examined  in this work for modeling LBS glasses using DeepMD.
The glass structures obtained by this MLP possessed four-fold coordinated boron ($^4$B) confirmed well with the experimental data
and abundance of three-membered rings.
The models were energetically more stable compared with those constructed with a functional force-field
even though both the models included reasonable $^4$B.
The results confirmed MLP to be superior to model the boron-containing glasses 
and address the inherent shortcomings of the AIMD and CMD.
This study also discusses some limitations of MLP for modeling glasses.

\end{abstract}

\newpage

\section{Introduction}

Boron containing silicate glasses are widely used for a variety of products such as 
tableware \cite{bouras2009}, optical fibers \cite{griscom1976}, display glass \cite{dejneka2019}, sealing glass \cite{lin2011}, bioactive glass \cite{hench2006,fu2010}, and nuclear waste glass \cite{jantzen1986}
because of their specific features such as high chemical durability, low thermal expansion coefficient, and high thermal shock resistance \cite{bengisu2016}.
Among them, lithium-containing borosilicate glasses are expected to play an important role
in all-solid-state battery technologies \cite{kluvanek2007,kim2011,deshpande2002,tsuda2018} because of their flexibility, which enables a
smooth connection between the electrolyte and electrode to suppress the interfacial resistance \cite{berkemeier2007,tatsumisago2013,wu2020}.
 
An extensive understanding of the atomistic view of the microstructure is indispensable to 
precisely preserve glass properties such as ion conductivity, mechanical response, and crack resistance.
Quantum calculation with density functional theory (DFT) is a promising method
to accurately obtain material structures in the nano-scale
because the computational cost of DFT calculations is relatively low compared with that of more precise approximations such as
the Moller–Plesset perturbation, coupled cluster calculations, and configuration interaction method.
Nevertheless, DFT calculation is still an expensive tool for modeling amorphous oxide glasses, mundanely
because many atoms should be considered to represent the variations of the amorphous structure 
while avoiding the artifact of the periodic boundary condition, which is often assumed to represent an infinite material structure
in computational modelings.
Besides, a significantly long cooling time is usually required to obtain well-equilibrated glass structures.
Therefore, even the DFT calculations often only qualitatively reproduce the glass microstructures; 
i.e., the number of tetrahedral BO$_4$ units and planar three membered rings, such as a boroxol ring, are underestimated \cite{ferlat2008,ohkubo2021}. 

Although, a possible alternative method to obtain adequately large and well-equilibrated glass structures 
is the classical molecular dynamics (CMD) simulation,
CMD simulations often fail to reproduce borosilicate glass structures.
Contrary to silicate glasses, 
borosilicate glasses including modifiers are difficult to be modeled accurately with simple functional force-fields
because boron atoms flexibly vary the oxygen coordination number, three-fold ($^3$B) or four-fold coordinated ($^4$B) \cite{zhong1988,stebbins1996,hubert2014}. 
To accurately model sodium borosilicate glasses using CMD simulations,
several force-fields were extended by employing composition-dependent parameters \cite{kieu2011,deng2016,deng2019,sund2020}. 
Analogously, a composition dependent force-field for lithium borosilicate glass was also proposed
by modifying the Buckingham type force-field \cite{urata2021a,urata2021b,urata2022a}, whose parameters were optimized to reproduce force and energy evaluated by DFT calculations \cite{urata2022b}.
The empirical force-field for LBS glasses is abbreviated as FMP-LBS, hereafter \cite{urata2022b}.
These extended empirical force-fields can reproduce $^3$B/$^4$B ratios in alkaline borosilicate glasses; however,
it is known that planar three membered rings such as boroxol rings are seldom in  
glass structures modeled by force-fields with two-body interactions \cite{takada1995}.

A recent progressive advancement of machine-learning (ML) technologies provides us with another approach to investigate
glass models more precisely \cite{kocer2021}.
In addition to relatively shallow layered neural networks with descriptors defined by atom-centered symmetry functions \cite{behler2007}, 
deep-learning neural networks \cite{schutt2017,takamoto2022a,takamoto2022b}, and graph-based network models \cite{gilmer2017,schutt2018} are extensively used to develop
interatomic potential models for simulating materials.
These ML potentials (MLPs) are computationally more expensive than the conventional functional force-fields,
whereas the computational cost of MLPs is substantially lower than that of DFT calculations.
Therefore, the CMD simulations enriched by MLPs would satisfy both the accuracy and computational cost for
modeling amorphous glass structures.
Indeed, MLPs have been applied to model silica \cite{takamoto2022a} and borosilicate glasses \cite{urata2022a}, for instance. 
This study, therefore, examine the applicability of MLP to accurately model LBS glasses.

The rest of this paper is organized as follows.
The computational procedures to train the MLP and to model the LBS glasses are presented in Section 2. 
In Section 3, after examining the boron coordination change with glass compositions,
three-membered ring formation in the LBS glasses is observed. 
Dynamical property relating to the Li ion conductivity is also examined.
Section 4 summarizes the computational results and provides subjects to be studied further.

\section{Simulation Method}
Following our previous work \cite{urata2022a} on borosilicate glasses, this study employed DeepMD \cite{zhang2018a} for the architecture to construct the MLP.
DeepMD is composed of a local embedding network to describe an atomic environment and 
a subnetwork consisting the encoding and fitting neural networks to evaluate atomic energy \cite{zhang2018a,zhang2018b,wang2018}.

Local environment of atom $i$ interacting with $N_i$ atoms within a cutoff distance $r_c$ 
is represented by a local embedding matrix ($g$) composed of $N_i \times M_1$ elements as
\begin{eqnarray}
(g^i)_{jk} = \Bigl( G \bigl( s (r_{ij}) \bigr) \Bigr)_k , 
\end{eqnarray}
where matrix $G$ represents the local embedding neural network dependent on the pair of atom types for atoms $i$ and $j$,
and $r_{ij}$ is the distance between two atoms.
The matrix $G$ outputs $M_1$ values from the single value $s$($r_{ij}$), which 
smoothly reaches to zero at the cutoff distance using the truncated function as, 
\begin{equation}
s(r_{ij}) = 
\begin{cases}
    {1 \over r_{ij} } &  r_{ij} < r_{cs} . \\ 
    {1 \over r_{ij} } \Bigl \{{1 \over 2} cos \bigl[\pi { r_{ij} - r_{cs} \over r_c - r_{cs} } \bigr]  \Bigr \}  & r_{cs} \le r_{ij} \le r_{c} . \\
    0,        & r_{ij} > r_c .
\end{cases}
\end{equation}
Accordingly, the relationship between atoms $i$  and $j$ is defined as, 
\begin{eqnarray}
\widetilde{R}: \{ s(r_{ij}), \hat{x}_{ij}, \hat{y}_{ij}, \hat{z}_{ij} \} ,
\end{eqnarray}
where $r_{ij} = ||(x_{ij}, y_{ij}, z_{ij})||$, and 
\begin{eqnarray}
\hat{x}_{ij} = { s(r_{ij}) x_{ij} \over r_{ij} } .
\end{eqnarray}
The parameters $r_{cs}$ and $r_c$ were set to 5.8 and 6.0 \AA, respectively.

Using the local embedding matrix $g$, 
the encoded feature matrix $D$ is defined as
\begin{eqnarray}
D^i = (g^{i1})^T \hat{R}^i (\hat{R}^{i})^T g^{i2} ,
\end{eqnarray}
where $g^{i1}$ and $g^{i2}$ are the coordinate and axis filters, respectively.
Assuming $g^{i1}$ is $g^i$ and $g^{i2}$ is a part of $g^i$ until $M_2$,
$D^i$ is a matrix with a size of $M_1 \times M_2$.   
$\hat{R}^i (\hat{R}^{i})^T$ should be the symmetric matrix 
with translational and rotational invariance.
Then, atomic energy ($E_i$) is calculated by the sub-network composed of an encoding and fitting neural network
using the feature matrix $D_i$.
Optimal parameters of the network were obtained by an iterative training 
to minimize errors for force, energy, and virial calculations calculated by DFT. 
Readers can refer original and related works for more details on the algorithm.\cite{zhang2018a,zhang2018b,wang2018,jia2020} 

In total 9749 datasets were used for training the MLP, as summarized in Tables S1-3 of the supporting information (SI).
The training datasets included the crystalline structures of B$_2$O$_3$, Li$_2$O, Li$_2$Si$_2$O$_5$, Li$_2$SiO$_3$, and LiBSiO$_4$ 
obtained by $ab$ $initio$ MD (AIMD) simulations and
amorphous structures of silica, lithium silicate, lithium borate, and LBS glasses.
Most of the amorphous glass structures for the training data were calculated using CMD simulations with the interatomic potentials of FMP-LBS or Teter potential \cite{deng2019}, 
whereas the data with three LBS glasses were also obtained by conducting AIMD simulations.
These training data comprised data on the glass structures calculated at temperatures from 300 to 3500 K 
and deformed configurations to inform a variety of microstructures to the MLP.

All the DFT calculations were performed using the Vienna $ab$ $initio$ package (VASP) \cite{kresse1996a,kresse1996b} 
with the Perdew-Burke-Ernzerhof 
(PBE)  \cite{perdew1996} exchange-correlational functional for the generalized gradient approximation. 
The cutoff energy of the wave
function was 600 eV, and the energy conversion criterion was set to $1 \times 10^{-5}$ eV.
All the glass models and large crystal models were calculated at a single gamma point,
while AIMD simulations for the small crystal models were conducted with several K-points, as summarized in Table S1.

The parameter set for the DeepMD is summarized in Table S4 of the SI.
The CMD simulations with the FMP and Teter potential were conducted using the LAMMPS package \cite{plimpton1995}. 
The atom motions were integrated with 1 fs time step,
and temperature and pressure were controlled with a Nos\'{e}-Hoover thermostat \cite{nose1984} and barsotat \cite{tuckerman2006}, respectively.
The MLP-based CMD simulations were conducted using the ASE package \cite{bahn2002,larsen2017} under a canonical ensemble (NVT) with a time step of 1 fs.

\section{Simulation Results}
\subsection{Training Machine-Learning Potential}

Figure S1 of the SI shows mean and maximum of absolute errors (MAEs and MXEs, respectively) of energy and force, respectively, 
for the test datasets.
Additionally, parity plots of energy and force were visually shown in Figures S2 and S3, respectively,
for all the test datasets.
The test datasets were 10\% of the total dataset and not used for MLP training.
The training of MLP was sufficiently converged until a million iterations, as shown in Figure S4.
The overall MAE and MXE for energy were 0.0049 and 0.0946 eV/atom, respectively.
Those for force were 0.203 and 6.71 eV/\AA, respectively.
The MAEs of the FMP for silicate \cite{urata2021a}, aluminosilicate \cite{urata2021b}, and borosilicate glasses \cite{urata2022a} 
were approximately 0.5 eV/\AA~ and 0.01 eV/atom for force and energy, respectively.
These results confirmed the superior accuracy of the MLP compared to FMP-LBS
owing to the deep-learning architecture.

The transferability of the MLP was examined by evaluating the errors for 256 additional  test datasets, as shown in Figure 1.
The datasets for validation included three glass models such as LBS-20-60-20, LBS-40-40-20, and LBS-10-50-40 consisting of approximately 300 atoms. 
(Herein, the numbers in the abbreviations are mole ratios of SiO$_2$, B$_2$O$_3$, and Li$_2$O from the left, as shown in Table 1.) 
The glass structures were obtained upon the simulations cooling from 2000 K to 300 K using a temporarily optimized MLP.
The MAEs of force were 0.261, 0.243, and 0.218 eV/\AA~ for LBS-20-60-20, LBS-40-40-20, and LBS-10-50-40, respectively.
Notably, the MLP could also estimate a significantly large force in a glass model of LBS-40-40-20.
These test results ensured that the optimized MLP was adequately accurate for modeling LBS glasses,
although a consistent energy shift was observed for energy. 
Similarly, some test datasets exhibited consistent shift of energy from the DFT data, as shown in Figure S2.
In contrast, forces were reproduced well without shifting for any test dataset, as shown in Figure S3.
It is thus expected that the training of the MLP put priority on force rather than energy, 
and the consistent energy shift might be improved by tuning the weights of energy and force in an assessment function for 
evaluating the total error of MLP.

\begin{figure}[htb]
\begin{center}
\includegraphics[width=6.0in]{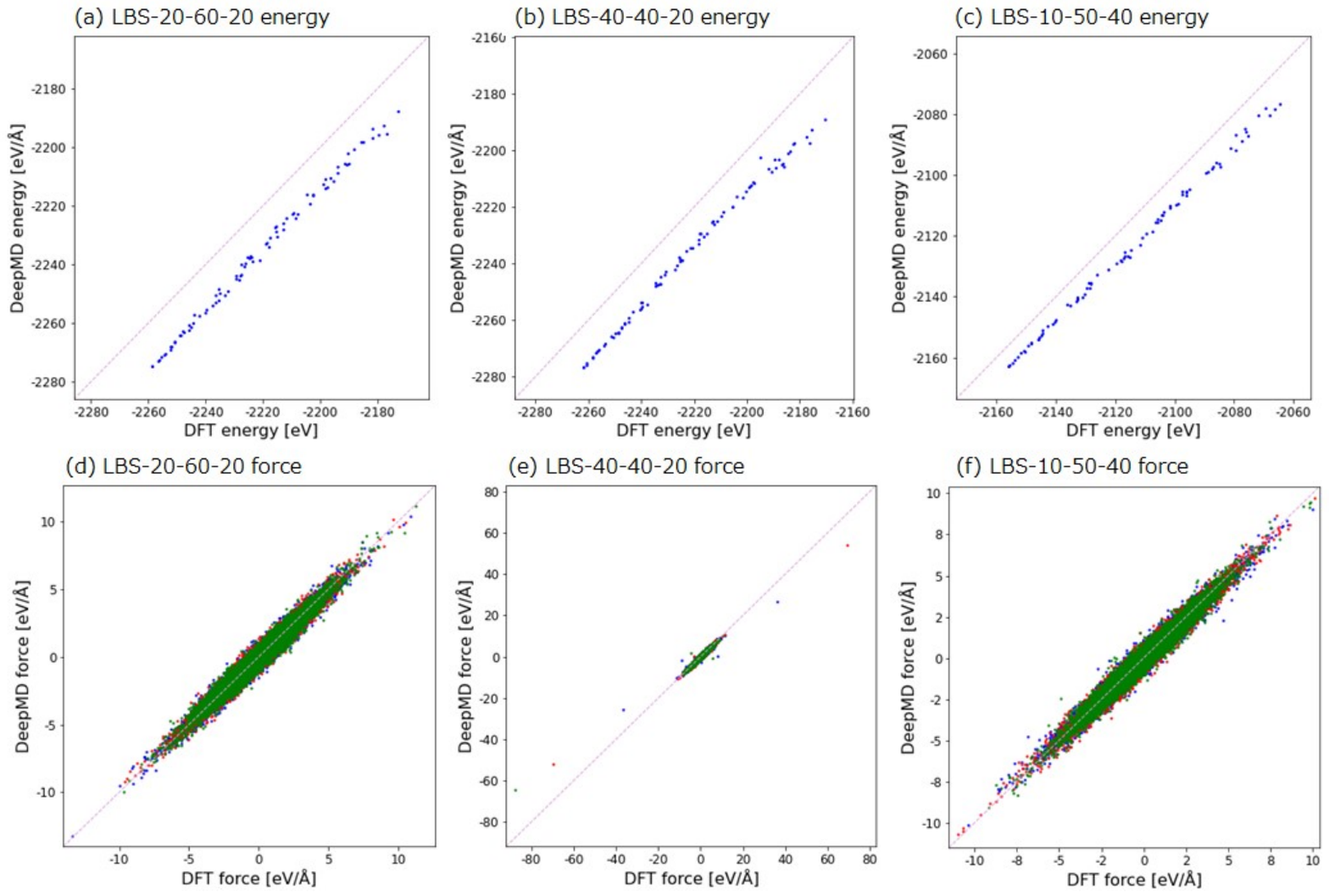}
\end{center}
\caption{Prediction results of potential energy and force using DeepMD for the additional test datasets.
(a)-(c) are energies for LBS-20-60-20, LBS-40-40-20, and LBS-10-50-40 glasses, respectively.
(d)-(f) are forces for LBS-20-60-20, LBS-40-40-20, and LBS-10-50-40 glasses, respectively
}
\label{fig1}
\end{figure}

\subsection{Boron coordination}

Using the MLP, eleven types of LBS glasses were modeled to compare with the experimental data on $^3$B/$^4$B ratio \cite{tsuda2018}, as summarized in Table 1.
First, we examined the small models, which are composed of approximately 300 atoms.
The atom numbers for modeling each glass are summarized in Table S5 of the SI.
The initial configurations were obtained using FMP-LBS via the usual melt-quenching method.
In the melt-quench simulations, a random configuration was melted at 3500 K for 500 ps.
It was then cooled down to 300 K
at a cooling rate of 1 K/ps followed by 300 ps simulations at 300 K.
During the CMD simulations with FMP-LBS, the MD cell volume was fixed according to the experimentally measured density.

Next, the force-field was switched to MLP.
The final configurations obtained by FMP-LBS were heated up to a high temperature at a heating rate of 10 K/ps,
then the atom configurations were mixed well at high temperature for 200 ps to remove the 
memory of the configurations constructed with FMP-LBS.
The maximum temperatures to melt the glass models were dependent on the glass compositions and ranged from 1500 to 2000 K (see Table S5 of the SI). 
To examine the effect of cooling rate on the glass structures, two cooling processes were investigated:
one process monotonically cooled the glass model at a cooling rate of 10 K/ps from the melting temperature down to 300 K,
whereas the other one maintained the temperature at 1000 K for 200 ps following the initial cooling process.
The glass models were then cooled down to 300 K at a cooling rate of 2 K/ps.
All models were equilibrated at 300 K for 200 ps, and the last 100 ps configurations were used for the structural analyses.

\begin{table}[htb]
\caption{LBS glasses studied by MD simulations with the MLP, and their experimentally observed four-fold coordinated boron ($^4$B) [\%] and density [g/cm$^3$].
The unit of glass composition is mol\%.
}
\centering
\small
\begin{tabular}{ l c c c c c c c }
\hline\noalign{\smallskip}
Glass name & SiO$_2$ & B$_2$O$_3$ & Li$_2$O & [SiO$_2$]/[B$_2$O$_3$] & [Li$_2$O]/[B$_2$O$_3$] & $^4$B & Density \\
\noalign{\smallskip}\hline\noalign{\smallskip}
LBS-20-60-20 & 20 & 60 & 20 & 0.33 & 0.33 & 36.7 & 2.21 \\
LBS-40-40-20 & 40 & 40 & 20 & 1.00 & 0.50 & 46.4 & 2.26 \\
LBS-10-50-40 & 10 & 50 & 40 & 0.20 & 0.80 & 41.6 & 2.30 \\
LBS-20-40-40 & 20 & 40 & 40 & 0.50 & 1.00 & 50.7 & 2.31 \\
LBS-30-30-40 & 30 & 30 & 40 & 1.00 & 1.33 & 54.4 & 2.31 \\
LBS-40-20-40 & 40 & 20 & 40 & 2.00 & 2.00 & 56.5 & 2.34 \\
LBS-50-10-40 & 50 & 10 & 40 & 5.00 & 4.00 & 56.8 & 2.32 \\
LBS-8-32-60  & 8 & 32 & 60 & 0.25 & 1.88 & 26.3 & 2.17 \\
LBS-15-25-60 & 15 & 25 & 60 & 0.60 & 2.40 & 22.0 & 2.20 \\
LBS-10-20-70 & 10 & 20 & 70 & 0.50 & 3.50 & 1.2 & 2.10 \\
LBS-5-35-60 & 5 & 35 & 60 & 0.14 & 1.71 & 23.5 & 2.17 \\
\noalign{\smallskip}\hline\noalign{\smallskip}
\hline\noalign{\smallskip}
\end{tabular}
\label{tab1}   
\end{table}

Figure 2(a) shows the ratio of $^4$B obtained by MLP.
Interestingly, the overall trend of the experimentally measured  $^4$B ratio is 
reproduced by the MLP with a few exceptions,
even though there are no artificial parameters dependent on the glass compositions in MLP
contrary to FMP-LBS and other empirical force-fields.
Meanwhile, the MLP underestimated the $^4$B ratio when the LBS glasses were quenched at a cooling rate of 10 K/ps.
Upon applying a slower quenching with 2 K/ps after the equilibration at 1000 K,
most of the LBS glasses exhibited higher $^4$B ratio, indicating that a longer relaxation led the glass configurations 
closer to the experimentally observed ones.  
To verify the reproducibility of the CMD simulations with MLP, we conducted three simulations by varying the
equilibration time at 1000 K as 200, 400, and 600 ps,
and the standard deviations of the three simulations were drawn as error bars in Figure 2(a).
The adequately small deviations of the $^4$B ratio confirmed the reproducibility of the MLP-based CMD simulations. 
Furthermore, the configurations after energy minimization with MLP were found to possess a $^4$B ratio closer to that of the experimental data,
implying that a further slower quenching would provide more accurate structures conforming with the experimental measurements.

It could be noted that the $^4$B ratio of LBS-50-10-40 was evidently underestimated, implying that
there was further room to improve the accuracy of MLP.
Additionally, the drawback of MLP could be seen upon analyzing the coordination number of the silicon atoms.
Silicon usually forms a tetrahedral four-fold coordinated structure in silicate glasses, but the glass models constructed by MLP
included a certain amount of five-fold coordinated silicon ($^5$Si) atoms.
These defect-like structures may not exist much in the experimental glass samples.
To confirm the stability of the glass structures obtained by MLP, further energy minimization was conducted by the DFT calculations with PBE
from the configurations optimized with MLP.
The $^4$B and $^5$Si ratios were almost maintained even after the energy minimization, as shown in Figure 2(b), confirming that the
glass structures obtained by MLP were at least in a local minimum of the energy landscape of the glass structures.
Since the borosilicate glasses obtained with DeepMD in our previous study did not contain $^5$Si \cite{urata2022b},
training more data on the structures wherein lithium and silicon are involved is expected to improve the MLP for modeling the LBS glasses.
Accordingly, this study focused on the  boron coordination and the three-membered rings (3MR) composed of boron atoms. 
 
\begin{figure}[htb]
\begin{center}
\includegraphics[width=6.0in]{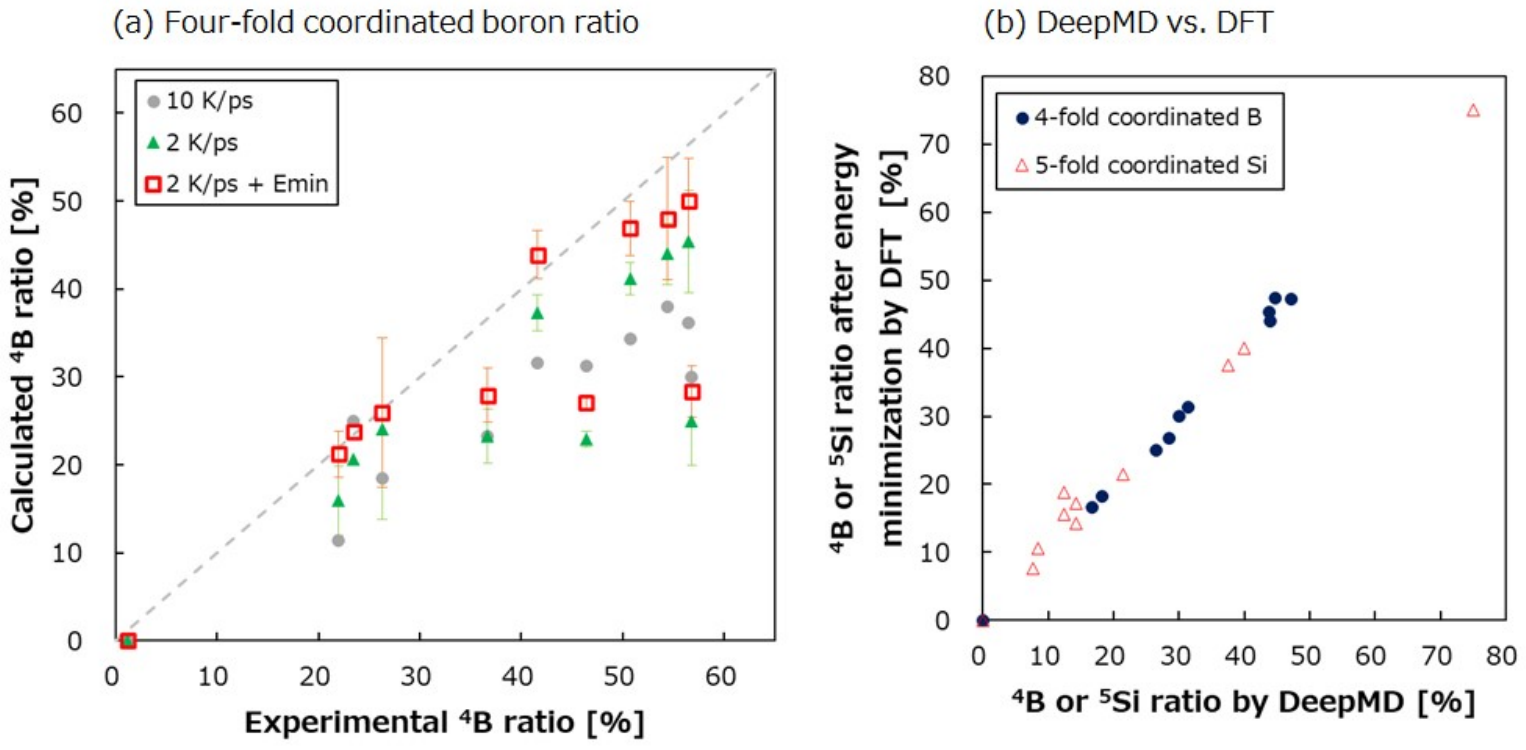}
\end{center}
\caption{(a) Comparison of four-fold coordinated boron ratios ($^4$B) between experiments and glass models with approximately 300 atoms. 
Error bars shows standard deviations of three independent simulations.
"2 K/ps + Emin" in the legend represents structures after energy minimization by the MLP.
(b) Comparisons between glass structures obtained by MD simulations with the MLP and energy minimization by DFT
on four-fold coordinated boron and five-fold coordinated silicon ratios.
}
\label{fig2}
\end{figure}

To validate the model size dependency, the larger models composed of approximately 1000 atoms (see Table S5 of the SI) 
were also examined, as shown in Figure 3.
The overall trend herein is analogous to the case of smaller models with 300 atoms, whereas
several differences are found.
For instance, the underestimation of the $^4$B ratio for the LBS-50-10-40 glass was well remedied, 
while the $^4$B ratio of the LBS-15-25-60 glass was overestimated contrary to the smaller model.
These inconsistencies indicate that the more training data may be necessary to cover the substantially large configurational space in
amorphous structures contrary to crystalline materials, even though the constructed MLP could
reproduce the $^3$B/$^4$B ratio in a non-empirical manner.
Furthermore, a benchmark test of the other types of MLPs would be valuable
to find the best architecture for modeling amorphous glasses as a subsequent work.

\begin{figure}[htb]
\begin{center}
\includegraphics[width=6.0in]{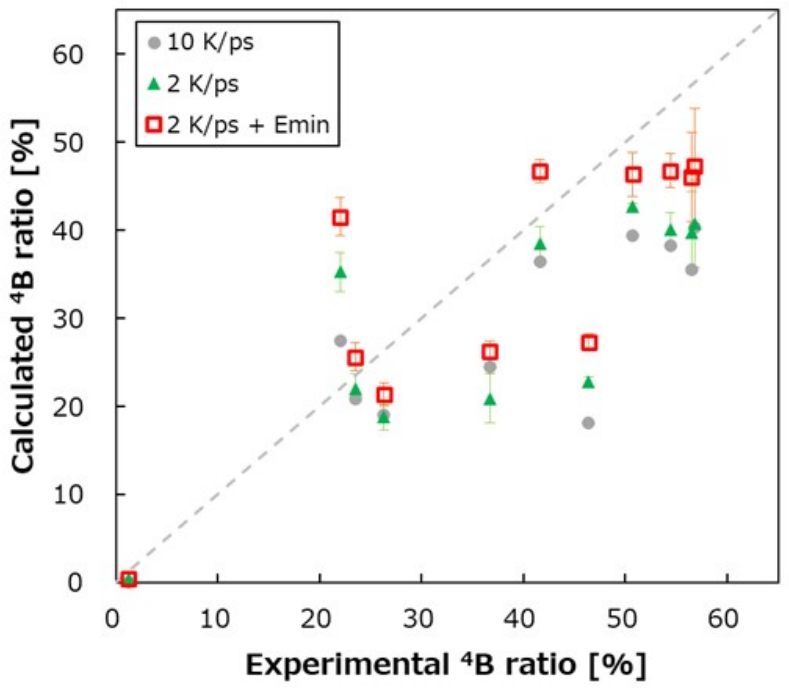}
\end{center}
\caption{Comparison of four-fold coordinated boron ratios ($^4$B) between experiments and glass models with approximately 1000 atoms. 
Error bars show standard deviations of three independent simulations.
"2 K/ps + Emin" in the legend represents structures after energy minimization by the MLP.
}
\label{fig3}
\end{figure}

\subsection{Three-membered ring}

The structures of 3MR, which is composed of B-O bonds, formed in the smaller and larger LBS glass models 
were analyzed using R.I.N.G.S. package \cite{le2010}, as shown in Figure 4(a).
The LBS-glasses with lower R = [Li$_2$O]/[B$_2$O$_3$] apparently possess  3MRs,
whereas the glasses comprising more Li$_2$O possess less 3MRs due to the limited connectivity among the network formers.
The typical atomistic configurations comprising the 3MRs with only B-O bonds are shown in Figure 4(b)-(e)
wherein the 3MRs are highlighted.
Additionally, the 3MRs were categorized into 4 types, 
which were defined by the ratio of $^3$B and $^4$B involved in each 3MR, as shown in Figure S5.
Specifically, the 3MR comprises three $^3$B is so-called as boroxol ring.
LBS-20-60-20, LBS-40-40-20, LBS-10-50-40, LBS-8-32-60 apparently possessed the boroxol rings, 
and the ratio of boroxol ring was related to the population of $^3$B.

For comparison of glass models obtained by MLP and FMP-LBS, 
Figure S6 shows the structure of LBS-20-60-20 glass modeled with 10001 atoms using FMP-LBS at 300 K
and the size distribution of rings composed of boron and oxygen.
The microsturcture includes only five 3MRs even in the ten times larger model, implying that FMP-LBS 
generates seldom 3MRs compared with MLP.
This is a remarkable difference in the glass models obtained by MLP and FMP-LBS.

Figure 5 compares energies evaluated by the DFT calculations for the glass structures constructed by FMP-LBS and MLP
for both smaller and larger models.
The simulation condition of the DFT calculations was same as that of the calculations on the training data.  
Additionally, the structures relaxed by energy-minimization by the DFT calculations were compared for the smaller models with 300 atoms. 
The energies of the glass models obtained with MLP were evidently lower than those of the glass structures 
constructed by FMP-LBS.
These results confirmed that MLP provided more appropriate glass structures compared to FMP-LBS
even though both the force-fields reproduced reasonable $^4$B ratio.
The cooling rate effect on the energy was subtle but a slower quenching generally provided more stable glass structures.
The energy minimization by DFT calculations further led the glass models to more stable structures, implying that 
the local structural distortions were remedied,
whereas the overall structural features were maintained since the $^4$B ratios didn't change much, as shown in Figure 2(b).    

These results demonstrated that MLP is superior to reproduce 3MRs compared with FMP-LBS; however, 
the numbers of 3MRs generated by MLP in the LBS-glass models were still less than those evaluated by NMR measurements \cite{ohkubo2021,du2003site}.
The NMR measurements distinguished the $^3$B atoms as $^3$B$_{ring}$, which constitutes a 3MR, and the others ($^3$B$_{nonring}$).
The ratio of $^3$B$_{ring}$ against overall $^3$B generated by MLP is compared with those analyzed by NMR  and AIMD simulations in Table 2.
For this comparison, 3MRs composed of Si-O and B-O bonds were also considered in the ring analysis.
According to the NMR experiments, $^3$B$_{ring}$ exists more than 50 \% in any LBS glass \cite{ohkubo2021,du2003site}, 
while those estimated by MLP were only 24.5 \% at maximum.
The results with MLP is almost comparable with AIMD simulation \cite{ohkubo2021}, even though the MLP enabled the slower quenching simulations. 
As observed in the NMR experiment on sodium borosilicate glasses, the $^3$B$_{ring}$ increased by longer annealing at lower temperature.
Therefore, it is expected that further slow quenching is necessary to construct 3MR more.
Furthermore, Du and Stebbins reported that the LBS glass exhibited great chemical heterogeneity \cite{du2003}.
It may be another possible reason for the underestimation of  $^3$B$_{ring}$, and
larger simulation model might be required in order to study the heterogeneous glass structure.

\begin{figure}[htb]
\begin{center}
\includegraphics[width=6.0in]{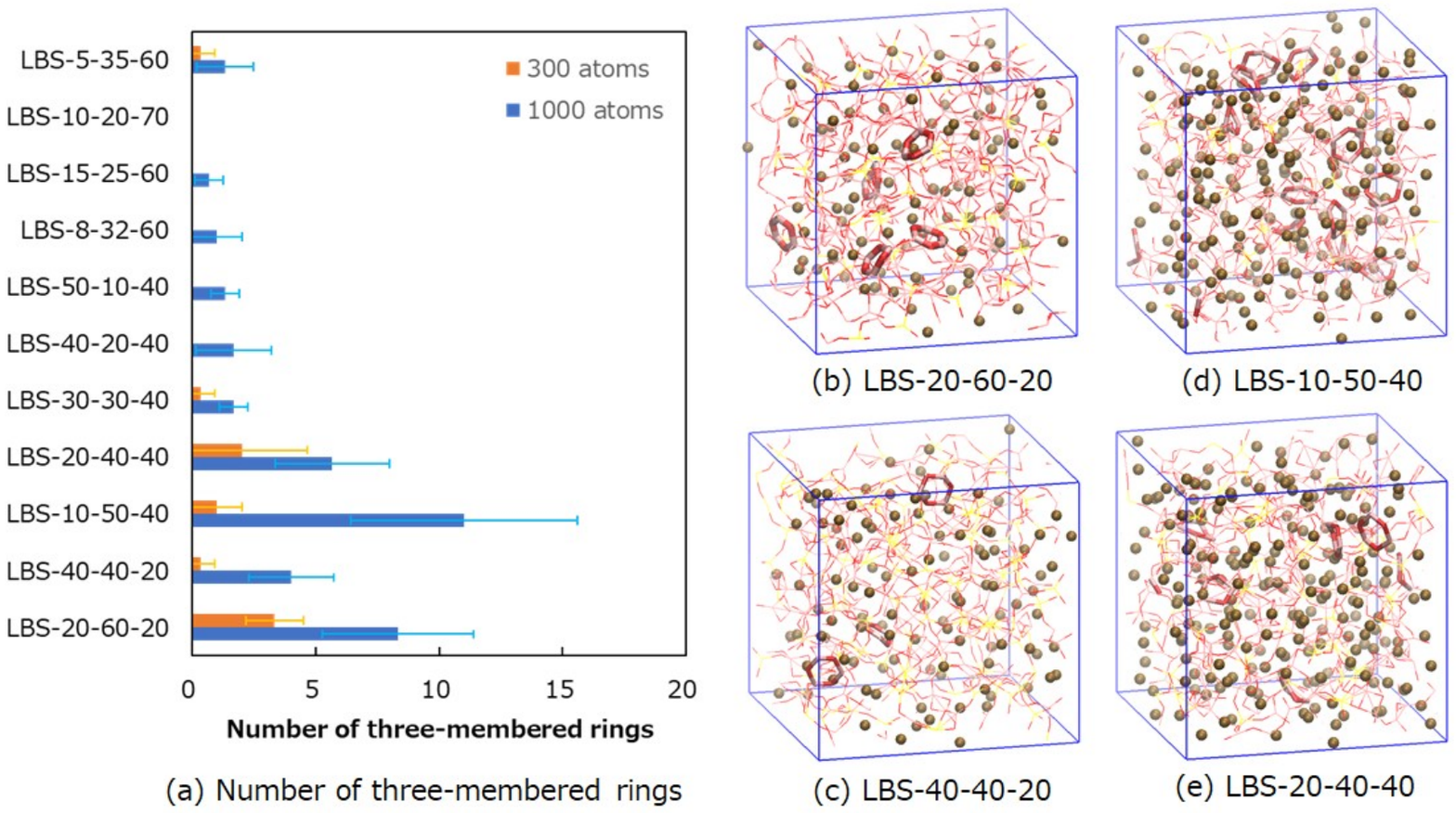}
\end{center}
\caption{(a) Number of three-membered rings, which is composed of B-O bonds, formed in the LBS glasses at 300 K.
The values are averaged for the three glass models cooled at a cooling rate of 2 K/ps.
The error bar indicates standard deviation of the three independent models.
(b)-(e) are typical glass structures comprising thee-membered rings. 
The rings are highlighted as bold bonds.
(Yellow) silicon, (red) oxygen, (pink) boron, and brown beads are Li ions.
}
\label{fig4}
\end{figure}

\begin{figure}[htb]
\begin{center}
\includegraphics[width=6.0in]{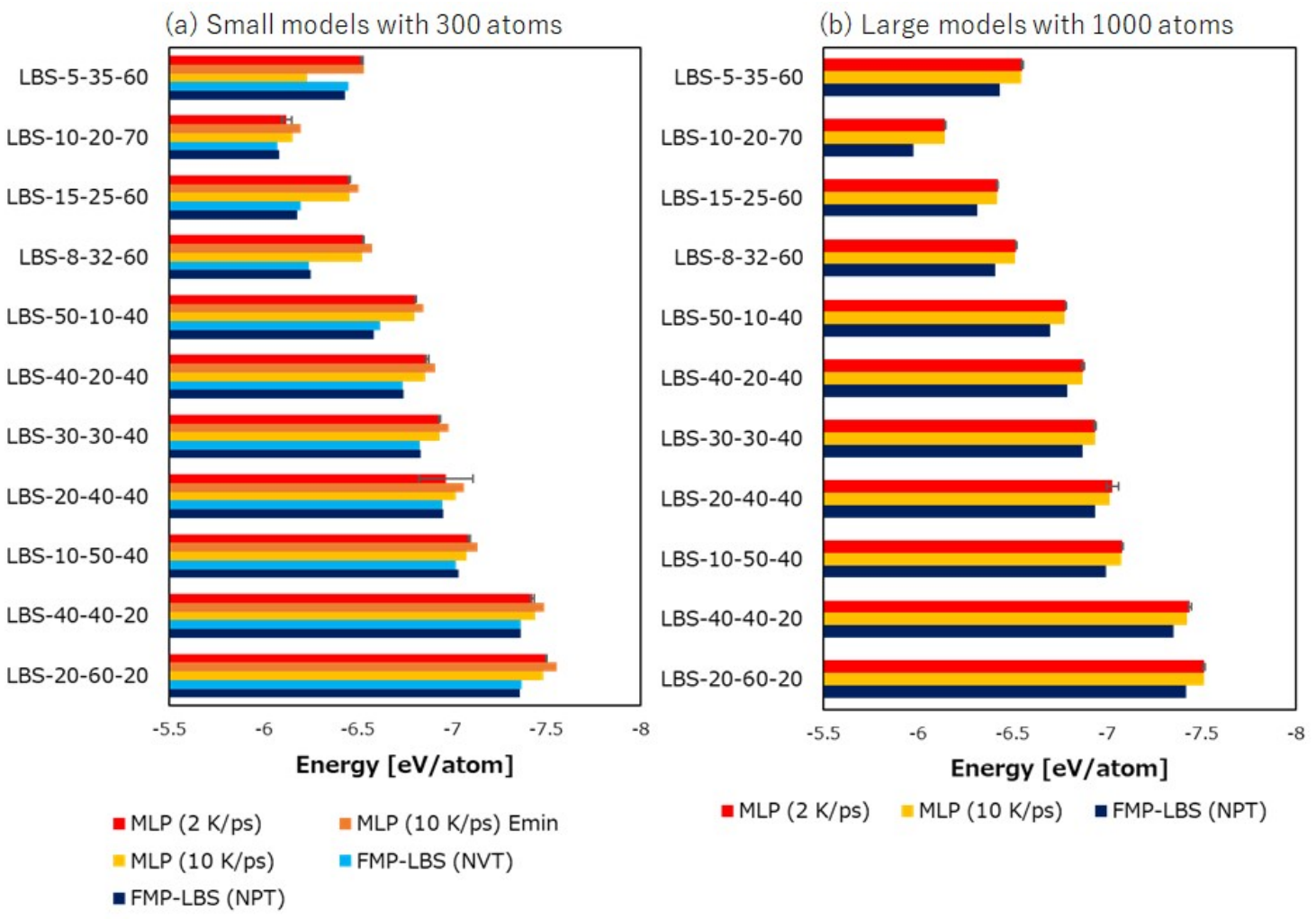}
\end{center}
\caption{DFT calculations of energies for the models obtained by MLP and FMP-LBS.
(a) Smaller models with approximately 300 atoms, (b) larger models with approximately 1000 atoms.
In the legend, values in the parenthesis represents cooling rate for CMD simulations with FMP, while
those for FMP-LBS mean ensemble used for quenching simulation at a cooling rate of 1 K/ps.
"Emin" means results by energy minimization with DFT caculations. 
}
\label{fig5}
\end{figure}

\begin{table}[htb]
\caption{
Comparisons among CMD simulations with MLP, AIMD simulations \cite{ohkubo2021}, and NMR measurements \cite{ohkubo2021,du2003site}
on the ratio of three-fold coordinated boron in three-membered ring structures ($^3$B$_{ring}$) against overall $^3$B in [\%].
}
\centering
\small
\begin{tabular}{ l c c c }
\hline\noalign{\smallskip}
Glass model & MLP  &  AIMD \cite{ohkubo2021} & NMR \\
\noalign{\smallskip}\hline\noalign{\smallskip}
LBS-20-60-20 & 15.4 $\pm$ 4.5 &    &  \\
LBS-40-40-20 & 18.6 $\pm$  5.9 &    &   \\
LBS-10-50-40 & 23.1 $\pm$  9.3 & 29.0    & 85.7 ~\cite{ohkubo2021} \\
LBS-20-40-40 & 24.5 $\pm$  2.1 &    &   \\
LBS-30-30-40 & 19.0 $\pm$  3.6 & 17.2 &  76.9  ~\cite{ohkubo2021}\\
LBS-40-20-40 & 22.2 $\pm$  10.1 &    &   \\
LBS-50-10-40 & 20.5 $\pm$  5.8 &    &   \\
LBS-8-32-60 & 5.0 $\pm$  1.0 &    &   \\
LBS-15-25-60 & 6.6 $\pm$  4.6 & 3.6  & 56.1 ~\cite{ohkubo2021}\\
LBS-10-20-70 & 0.0  &    &   \\
LBS-5-35-60 & 3.7 $\pm$  2.4 &    &   \\
LBS-53.3-26.7-20 & --  &    & 67.6 ~\cite{du2003site}  \\
\noalign{\smallskip}\hline\noalign{\smallskip}
\hline\noalign{\smallskip}
\end{tabular}
\label{tab2}   
\end{table}

Next, we examined the formation process of the 3MRs composed of boron and oxygen during quenching at a cooling rate of 10 K/ps
in the larger glass models of LBS-20-60-20, LBS-40-40-20, LBS-10-50-40, and LBS-20-40-40 glasses,
which possess a relatively large number of 3MRs.
The glass configurations were extracted from the highest temperature to 300 K every 10 K during cooling, and
the temperature dependence on the number of 3MRs was analyzed, as shown in Figure 6.
The 3MRs increase with the temperature decrements for LBS-20-60-20, 
while several 3MRs already exist in the melted glass structures at high temperature.
For the other three glasses, which possess less rings compared to LBS-20-60-20,
the variation in the number of 3MRs is not monotonic, and sometimes, more rings were formed 
at a relatively high temperature compared to the glass models obtained at 300 K.
This result potentially indicates that each boron atom frequently changed the bridging boron atoms at high temperatures,
and eventually only the stable rings remained at lower temperatures.

To see the process of 3MRs formation, some of the atomic configurations of LBS-20-60-20 
are visualized in Figure 7.
Figure 7(e) shows the abundance ratio of the 3MRs, which exist at 300 K, at each temperature. 
80\% of the 3MRs were constructed until 800-900 K, and some of them were already formed at a higher temperature,
indicating that the 3MRs were initiated to form at various temperatures.
The apparent jump between 900 and 1000 K may relate to the glass transition temperature (T$_g$),
wherein the glass structure was almost frozen.
This observation is consistent with the experimental study by Ike et al., 
which observed boroxol ring deformation above T$_g$ for a lithium borate glass using Brillouin scattering spectroscopy \cite{ike2006}.
In contrast, interestingly, some of the 3MRs still appeared and disappeared at lower temperatures.
Figures 7(a) and (b) show the rings formed at around 800 K.
In case (a), the 3MR was transformed from a four-membered ring, while
three neighboring three-fold-coordinated boron atoms formed the three-membered boroxol ring at 800 K in case (b).
As shown in Figure 7(c), the 3MR sometimes disappears during cooling
due to the instability of the ring structure.
Contrarily, once stable 3MRs form at high temperatures, they survive until low temperatures,
as shown in Figure 7(d).  

To confirm the glass transition temperatures of the models constructed with MLP,
heating simulations were conducted at a heating rate of 10 K/ps.
Here, three independent simulations were carried out starting from the equilibrated models at 300 K
after quenching at a cooling rate of 2 K/ps. 
According to the potential energy change with temperature, T$_g$ values were determined as a folding point of potential energy curve, as shown in Figure S7.
Consequently, it was confirmed that the T$_g$ of LBS-20-60-20 glass was 1122 $\pm$ 31 K,
which almost corresponds to the temperature at which the 3MRs increased, discontinuously.
Therefore, it is expected that 3MR might become more stable at temperature lower than T$_g$.
Furthermore, the T$_g$ values of the LBS glasses modeled by the MLP were found to correlate well with the experimental data \cite{tsuda2018}, as shown in Figure 8,
even though the MLP consistently overestimated T$_g$ as in the case of most of MD simulations \cite{urata2019}.
This good agreement confirmed the accuracy of the glass models obtained by the MLP. 

\begin{figure}[htb]
\begin{center}
\includegraphics[width=5.0in]{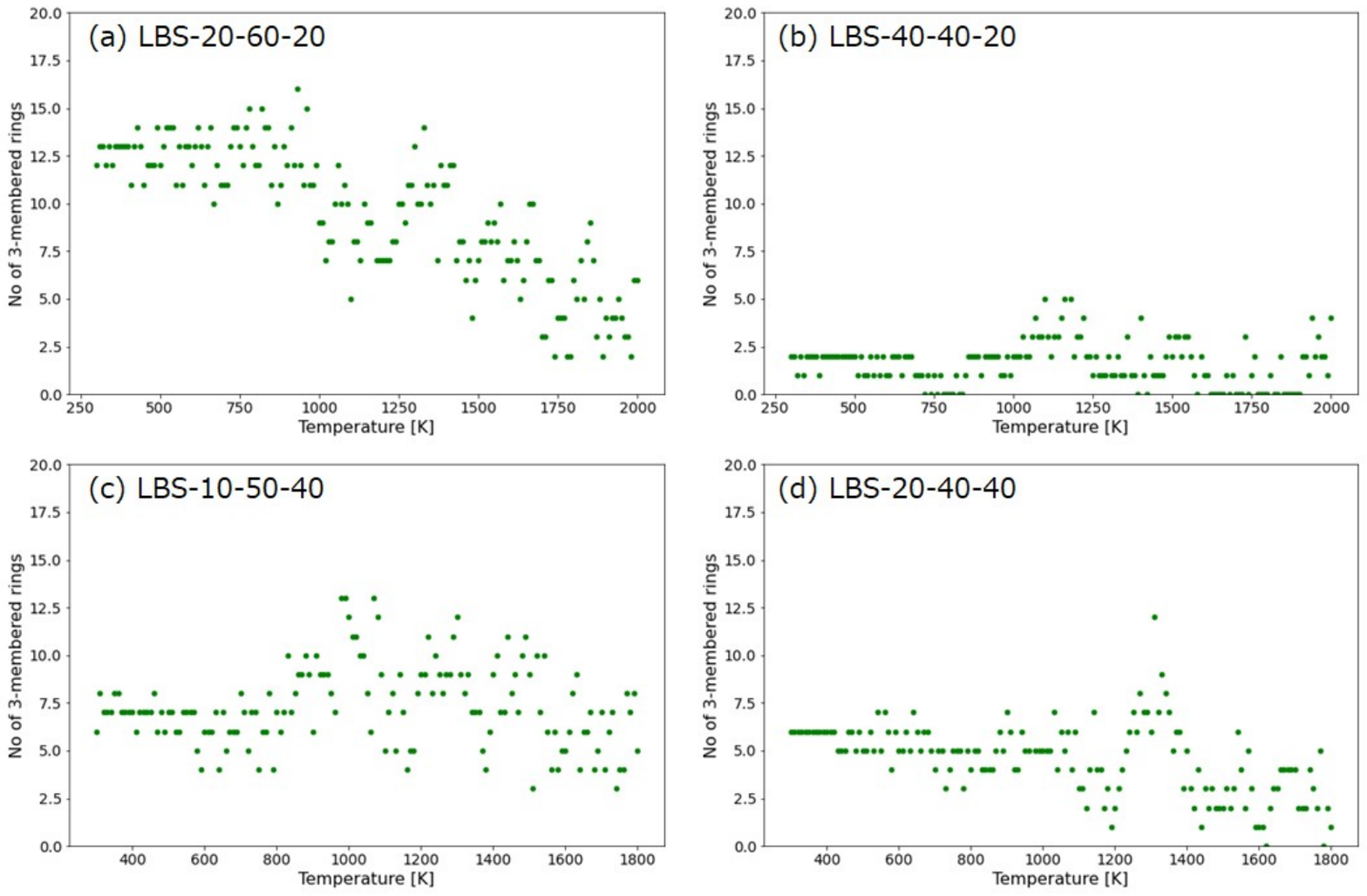}
\end{center}
\caption{Variations of number of the three-membered rings composed of three boron atoms for 
(a) LBS-20-60-20, (b) LBS-40-40-20, (c) LBS-10-50-40, and (d) LBS-20-40-20
while cooling from the melting temperature to 300 K. 
Results on the larger models (~1000 atoms) cooled at a cooling rate of 10 K/ps.
}
\label{fig6}
\end{figure}

\begin{figure}[htb]
\begin{center}
\includegraphics[width=6.0in]{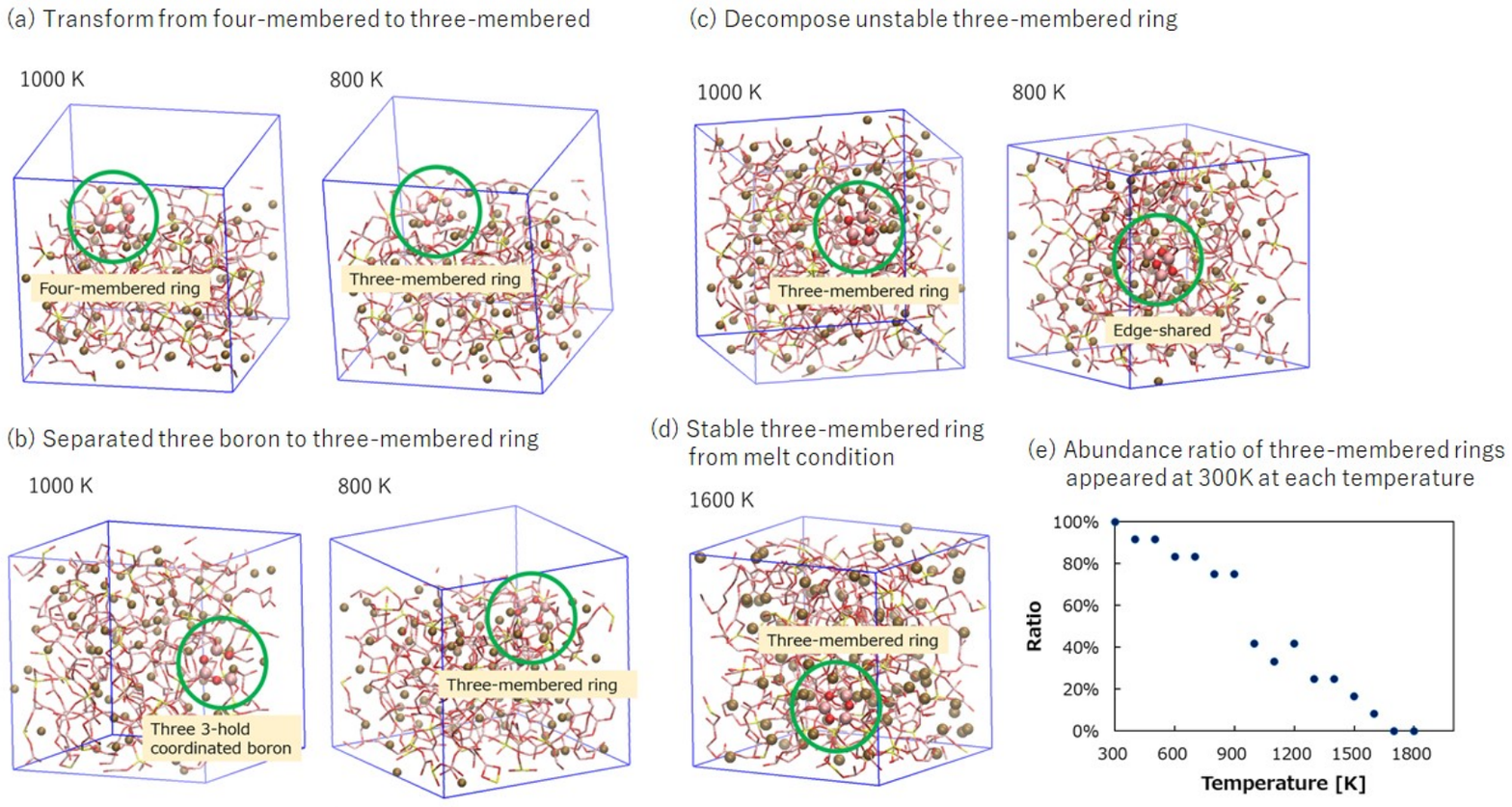}
\end{center}
\caption{Structural changes to form three-membered rings (a) from a four-membered ring and 
(b) separated three three-fold-coordinated boron atoms.
(c) Decomposition of a three-membered ring at lower temperature.
(d) A stable three-membered ring during cooling from 1600 K to 300 K.
(e) Abundance ratio of the three-membered rings appearing at 300 K at each temperature during cooling.
All the results are taken from LBS-20-60-20 glass model with 1000 atoms
obtained via quenching at 10 K/ps of cooling rate.
}
\label{fig7}
\end{figure}

\begin{figure}[htb]
\begin{center}
\includegraphics[width=4.5in]{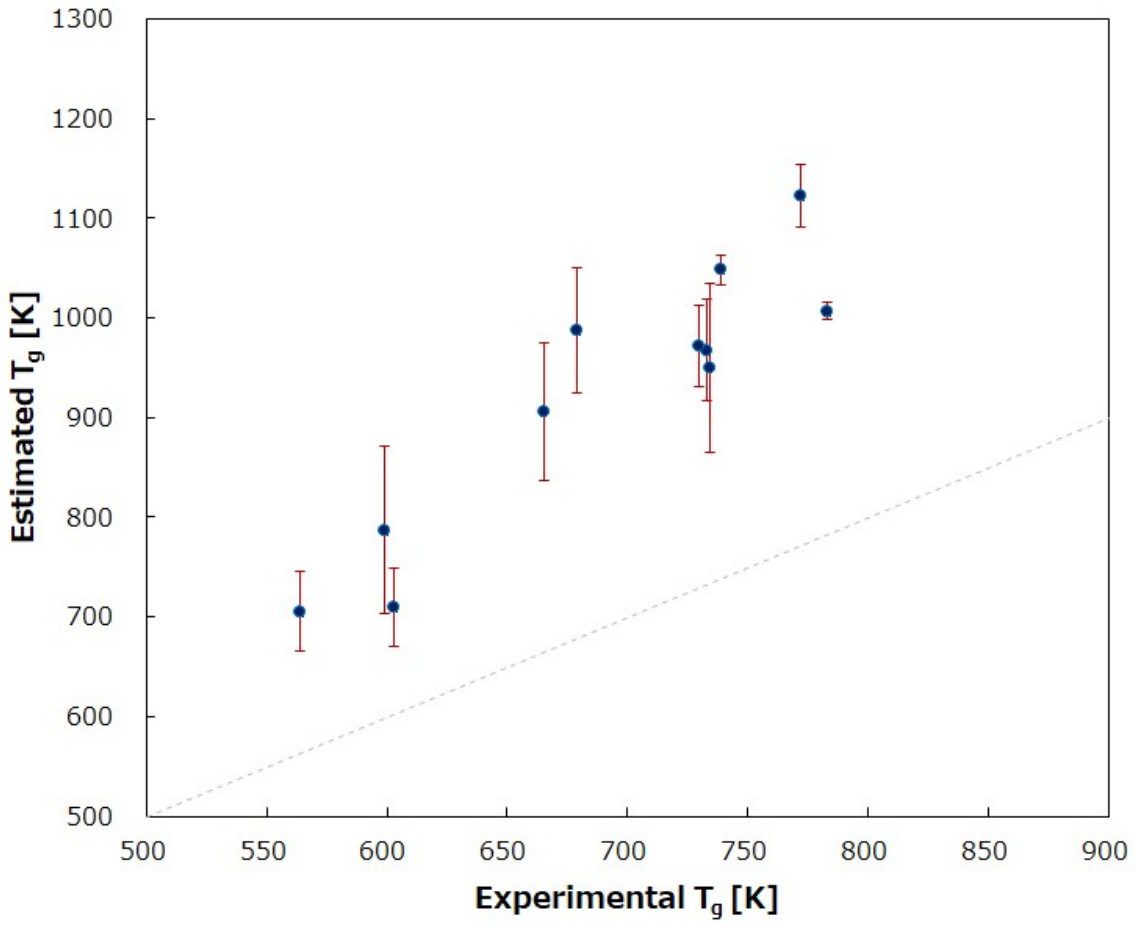}
\end{center}
\caption{Glass transition temperature (T$_g$) obtained by the MLP in comparison with experimental data \cite{tsuda2018}.
Error bars show standard deviations of three independent simulations.
}
\label{fig8}
\end{figure}

\subsection{Li ion conductivity}

To verify the accuracy of MLP for evaluating dynamic property, diffusion coefficients ($D$) of Li ions were calculated using the larger models with 1000 atoms.
The CMD simulations were conducted at 500 K for 300 ps with an NPT ensemble to clearly see the dynamics.
Then, the diffusion coefficients were evaluated from the linear relationship between time and mean square displacements (MSD) between 50 and 150 ps,
according to the atomic trajectories except for the initial 50 ps data.
\begin{eqnarray}
D = {MSD(t) \over 6t}.
\end{eqnarray}
Figure 8 compares $C_{Li} \times D$, where $C_{Li}$ is number density of Li ion, and experimental electric conductivity \cite{tsuda2018}.
Here, in total four models obtained with the cooling rates of 10 and 2 K/ps were used to calculate average and standard deviations of MSD.
Even though LBS-50-10-40 exhibits relatively large difference from experimental data,
overall trend of the electric conductivity is almost reproduced by the CMD simulations with MLP.
This result confirmed the applicability of MLP to explore Li-containing glasses exhibiting high electric conductivity.

Next, activation energy (E$_a$) of Li ion conductivity was estimated from the diffusion coefficients calculated at 500, 600, 700, and 900 K
for the LBS glasses comprising 40 mol\% of Li$_2$O (LBS-10-50-40, LBS-20-40-40, LBS-30-30-40, LBS-40-20-40, and LBS-50-10-40)
with assuming Arrhenius equation, 
$D = D_0 exp(-E_a/RT)$, and the results were summarized in Table 3.
The E$_a$ estimated by MLP was smaller for LBS glasses with higher B$_2$O$_3$ content, whereas it reached maximum at around 20-30 \% of B$_2$O$_3$.
This trend is somewhat similar to the experimental observation in \cite{maia2004}.  
However, the theoretical values were smaller than experimental E$_a$:
0.59 $\sim$ 0.71 eV for 0.4Li$_2$O$\cdot$0.6($x$B$_2$O$_3$(1-$x$)Si$_2$O$_4$) \{x = 0 $\sim$ 1 \} \cite{maia2004} and 
0.50 $\sim$ 0.56 eV for $x$Li$_2$O$\cdot$(1-$x$)(0.75B$_2$O$_3$0.25SiO$_2$) \{x = 0.5 $\sim$ 0.675 \} \cite{saetova2016}.
A possible reason is that this study applied the NVT ensemble with assuming experimental density at room temperature for any temperature condition.
Because the high density at higher temperature hinders ion dynamics, E$_a$ can be underestimated.

\begin{table}[htb]
\caption{The activation energy of Li ion conductivity for the LBS glasses with 40 mol\% of Li$_2$O.
}
\centering
\small
\begin{tabular}{ l c  }
\hline\noalign{\smallskip}
Glass model & Activation energy [eV] \\
\noalign{\smallskip}\hline\noalign{\smallskip}
LBS-10-50-40 & 0.37 $\pm$  0.02 \\
LBS-20-40-40 & 0.37 $\pm$  0.03 \\
LBS-30-30-40 & 0.40 $\pm$  0.05 \\
LBS-40-20-40 & 0.40 $\pm$  0.05 \\
LBS-50-10-40 & 0.38 $\pm$  0.02 \\
\noalign{\smallskip}\hline\noalign{\smallskip}
\hline\noalign{\smallskip}
\end{tabular}
\label{tab3}   
\end{table}

\begin{figure}[htb]
\begin{center}
\includegraphics[width=6.0in]{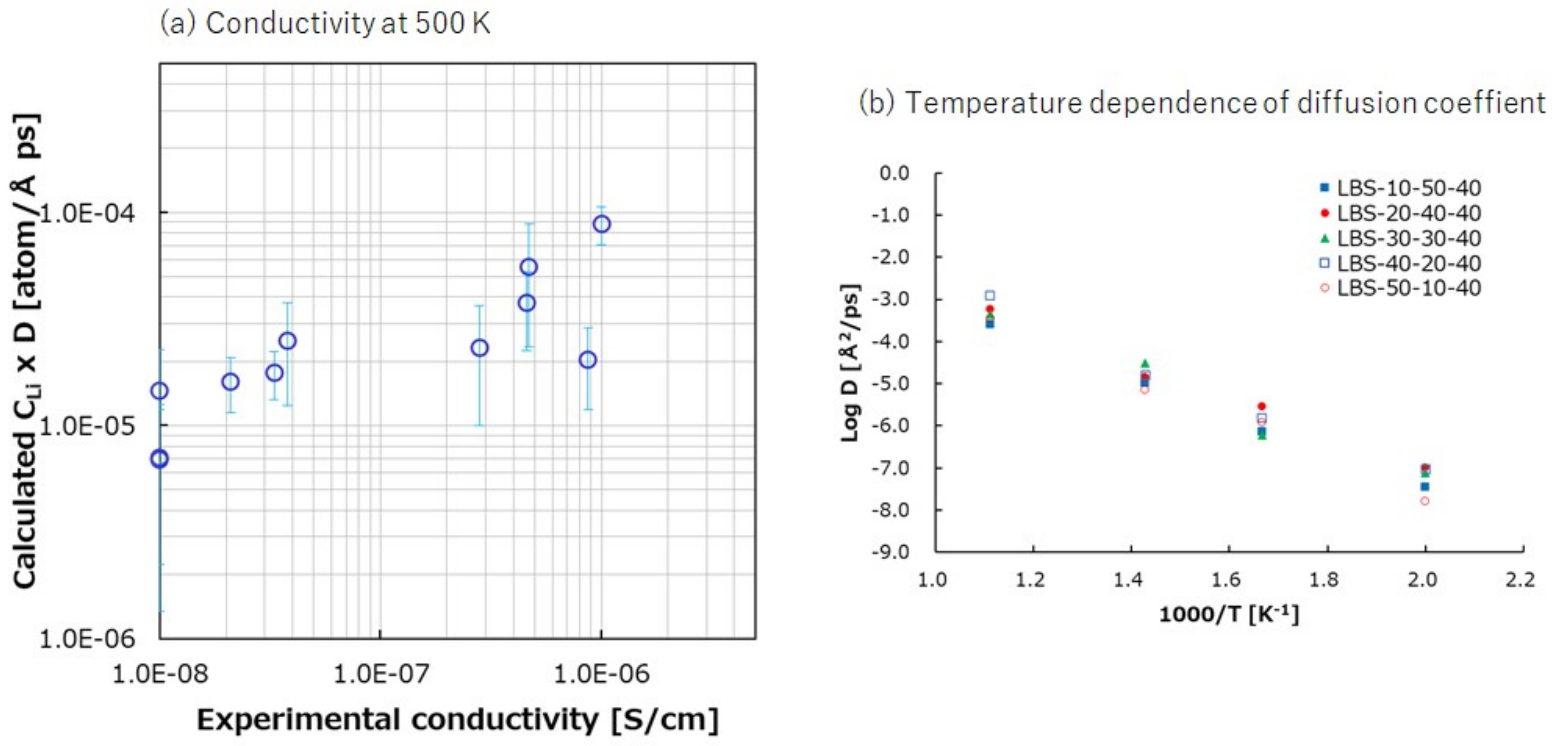}
\end{center}
\caption{(a) Comparison with experimental data on electric conductivity (Tsuda et al. \cite{tsuda2018}) 
and ion conductivity for LBS glasses modeled with approximately 1000 atoms.
Error bars represent standard deviation of four models constructed with the cooling rates of 10 and 2 K/ps.
The diffusion coefficients were evaluated at 500 K.
(b) An example of temperature dependence of diffusion coefficient of Li ion measured at 500, 600, 700, and 900 K
in order to determine the activation energy.
}
\label{fig9}
\end{figure}
 
\section{Conclusions}

In conclusion, this study demonstrated the superior accuracy of machine-learning potential to
model boron-containing glasses in comparison with conventional functional force-fields.
Furthermore, the lower computational cost of MLP compared to DFT calculations enabled us to 
apply a slower quenching for constructing larger glass models, essential to obtain well-equilibrated glass structures.
Consequently, three-fold and four-fold coordinated boron atoms were adequately formed in a variety of LBS glasses
without employing any artificial corrections contrary to the functional force-field. 
Additionally, three-membered rings composed of boron atoms were 
appropriately constructed, especially in high B$_2$O$_3$ content glasses.
The glass models constructed using MLP were energetically more stable than those obtained by FMP-LBS,
implying that a certain amount of three-membered rings should exist in such lithium borosilicate glasses.
The relatively large models with approximately 1000 atoms enabled us to observe a variety of formation processes
of the 3MRs during cooling.
One of the interesting findings was that the 3MRs were not formed at any specific temperature,
and the 3MRs sometimes disappeared if the structure was instable.
Once a stable 3MR structure appears even at a high temperature, it can survive during cooling. 

We also noticed several issues upon the application of MLP.
First, the $^3$B/$^4$B ratios for a few LBS glasses were inconsistent between the smaller and the larger models.
Second, some silicon atoms formed five-fold coordinated, which may be inexistent in experimental glass samples.
Third, at higher temperatures (i.e. 2000 K), some atoms closely approached  and finally overlapped each other,
as in the modeling of borosilicate glasses \cite{urata2022a}.
These undesired behaviors can be remedied by adding more training data to cover more configurational space
since amorphous materials are composed of more multiple microstructures contrary to crystalline materials.
Nevertheless, this study demonstrated unique possibility of MLP to model amorphous materials
comprising elements drastically changing their coordination and also planar ring structures 
formed by sp2 boron atoms.
Therefore,
it is worth comparing a variety of MLPs to find an appropriate approach to model amorphous materials using MLP
in subsequent studies.




\begin{suppinfo}
List of DFT data and hyperparameters for training MLP, and the fitting results.
Types of three-membered rings, ring size distribution, and fitting method to determine T$_g$.

\end{suppinfo}

\bibliographystyle{achemso}

\bibliography{LBS_DMD}

\newpage



\end{document}